\documentclass[english,twoside,a4paper,10pt]{article}

\usepackage[latin1]{inputenc}
\usepackage[T1]{fontenc}
\usepackage{amsmath}
\usepackage{amsfonts}
\usepackage{graphicx}
\usepackage{subfigure}
\usepackage{a4wide}
\usepackage{amssymb}
\usepackage{fancyhdr}
\usepackage{mathrsfs}
\usepackage[toc,page]{appendix}

\begin{document}

\title{\bf Inhomogeneous viscous fluids in FRW universe}
\author{ 
R. Myrzakulov$^1$\footnote{Email: rmyrzakulov@gmail.com; rmyrzakulov@csufresno.edu}\,,
L. Sebastiani$^{1}$\footnote{E-mail address:l.sebastiani@science.unitn.it
}\, and S. Zerbini$^2$\footnote{E-mail address:zerbini@science.unitn.it}\\
\\
\begin{small}
$^1$ Eurasian International Center for Theoretical Physics and  Department of General
\end{small}\\
\begin{small} 
Theoretical Physics, Eurasian National University, Astana 010008, Kazakhstan
\end{small}\\
\begin{small}
$^2$ Dipartimento di Fisica, Universit\`a di Trento, Italy and 
\end{small}\\
\begin{small}
 Gruppo Collegato di Trento, Istituto Nazionale di Fisica Nucleare, Sezione di Padova, Italy
\end{small}\\
}

\date{}

\maketitle


\begin{abstract}

We give a brief review of some aspects of
inhomogeneous viscous fluids in a flat Friedmann-Robertson-Walker Universe. In general, it is pointed out that several fluid models may bring 
the future Universe evolution to become singular, with the appearance of the so-called Big Rip scenario.
We investigate the effects of fluids coupled with dark matter in a de Sitter Universe, by considering several cases. Due to this coupling, 
the coincidence problem may be solved, and if the de Sitter solution is stable, the model is also protected against the Big Rip singularity.

\end{abstract}



\section{Introduction}

%
%

Since the discovery of the current cosmic acceleration~\cite{SN11,SN12,SN13,SN14,WMAP01,WMAP02,WMAP03,WMAP}, 
the dark energy issue has become one of the most interesting fields of research in modern cosmology. 
It is well known that there exist several descriptions of the current accelerated expansion of the Universe. 
The simplest one is the introduction of a small positive 
Cosmological Constant in the framework of General Relativity, so that one is dealing with a perfect fluid whose Equation of State parameter is $\omega=-1$, and this fluid is able to describe the current cosmic acceleration, but also, the 
use of other forms of fluid (phantom, quintessence, inhomogeneous fluids, \textit{etc}.), satisfying a suitable Equation of State is not excluded. 
On the other hand, the observed small value of the Cosmological Constant leads to several conceptual problems
(vacuum energy, coincidence problem, \textit{etc}.), so that in the last few years, several different approaches to the dark energy issue have been proposed. 
Among them, the modified theories of gravity~\cite{Review1,Review2,Review3,Review4,carroll1,carroll2,Carroll,Bamba,Hu,Hu-1,Hu-2,rev} represent an interesting extension of Einstein's theory, but also supersymmetry and string theories have been investigated.

In this short review, we will present some aspects of inhomogeneous viscous fluids in a flat Friedmann-Robertson-Walker Universe. 
The fluid representation of dark energy possesses many advantages. For example, besides the fact that we can still use the formalism of General Relativity by means of Friedmann equations, almost any modification to General Relativity can be encoded in a fluid-like form, so that the study of inhomogeneous viscous fluids is one of the easiest way to understand some general features of such a kind of alternative theory.
 
The paper is organized as follows. In Section {2}, we will introduce the formalism of inhomogeneous viscous fluids in a flat Friedmann-Robertson-Walker Universe, and we will show how it is possible to write a modification to gravity in the fluid-like form [specifically, we will consider $F(R)$-gravity]. Thus, we will consider an inhomogeneous fluid model that reproduces a viable cosmology, but 
that brings the future Universe evolution to become singular at a finite-future time. This is the Big Rip scenario. It 
is present in a large class of fluids and some other examples will be mentioned. In Section {3}, we will couple inhomogeneous viscous fluids with dark matter. The reason for such a coupling consists in the attempt to solve the coincidence problem in a de Sitter Universe, since the ratio between dark matter and fluid energy will depend on the coupling constant, almost independently from initial conditions. Furthermore, when the de Sitter solution is stable, it is possible to avoid the finite future-time singularities. 
In Section {4}, as a new result, we consider a different case of coupling between dark matter and fluid, and we will repeat the calculations of Section {3}. Conclusions are given in Section {5}.

We use units of $k_{\mathrm{B}} = c = \hbar = 1$ and denote the
gravitational constant, $G_N$, by $\kappa^2\equiv 8 \pi G_{N}$, such that
$G_{N}^{-1/2} =M_{\mathrm{Pl}}$, $M_{\mathrm{Pl}} =1.2 \times 10^{19}$ GeV being the Planck mass.


\section{Inhomogeneous Viscous Fluids, Modified Gravity and the Big Rip}

In this Section, 
we will briefly review the general form of inhomogeneous viscous fluids in \linebreak Friedmann-Robertson-Walker (FRW) space-time and we will see how such a kind of fluid may describe a viable dark energy cosmology with some different final scenarios. 
As already mentioned, the fact that the dark energy observed in our Universe has an Equation of State (EoS) parameter, $\omega$, very close to minus one, 
suggests that the introduction of a positive Cosmological Constant in Einstein's equation is the most realistic way to describe the current cosmic acceleration. However, other kinds of fluids (quintessence, phantom, inhomogeneous, viscous fluids) are not excluded, and the modified theories of gravity have a corresponding description in the fluid-like form. In fact, the equation of state of inhomogeneous viscous fluid in a flat Friedmann-Robertson-Walker space-time
described by the metric:
\begin{equation}
ds^{2}=-dt^{2}+a^{2}(t)d\mathbf{x}^{2}\,
\end{equation}
where $a(t)$ is the scale factor of the Universe, 
reads~\cite{fluidsOd2}:
\begin{equation}
p_{\mathrm{F}}=\omega(\rho_{\mathrm{F}})\rho_{\mathrm{F}}+B(\rho_{\mathrm{F}},a(t),H, \dot{H}...)\,\label{start}
\end{equation}
where $p_{\mathrm{F}}$ and $\rho_{\mathrm{F}}$ are the pressure and energy density of fluid, the EoS parameter, $\omega(\rho_{\mathrm{F}})$, may depend on the energy density and the bulk viscosity, $B(\rho_{\mathrm{F}},a(t),H, \dot{H}...)$, is a general function of the fluid energy density, the scale factor, the Hubble parameter and its derivatives. 

As we stated above, with this general form of time-dependent bulk viscosity, we can 
encode any modification to gravity in the fluid-like form. For example, in $F(R)$-gravity, the action is given by:
\begin{equation}
I=\int_{\mathcal{M}} d^4x\sqrt{-g}\left[
\frac{F(R)}{2\kappa^2}+\mathcal{L}^{\mathrm{(matter)}}\right]\,\label{actionF(R)}
\end{equation}
where $g$ is the determinant of the metric tensor, $g_{\mu\nu}$, $\mathcal{M}$ is the space-time manifold,
${\mathcal{L}}^{\mathrm{(matter)}}$ is the matter Lagrangian and 
$F(R)$ is a function of the Ricci scalar, $R$. The FRW equations of motion (EOMs) can be written in the form:
\begin{equation}
\rho_{\mathrm{eff}}=\frac{3}{\kappa^{2}}H^{2}\,,
\quad
p_{\mathrm{eff}}=-\frac{1}{\kappa^{2}} \left( 2\dot H+3H^{2} \right)\,
\end{equation}
where $\rho_{\mathrm{eff}}$ and $p_{\mathrm{eff}}$ are
the effective energy density and pressure of the modified gravity model:
\begin{eqnarray}
\rho_{\mathrm{eff}} &\equiv&
\rho_{\mathrm{m}} +
\frac{1}{2\kappa^{2}}
\left[ \left( F'(R) R-F(R) \right)-6H^2(F'(R)-1)
-6H\dot{F}'(R)
\right]\,
\label{rhoeffRG} \\ \nonumber\\
p_{\mathrm{eff}} &\equiv&
p_{\mathrm{m}} +
\frac{1}{2\kappa^{2}} \Bigl[
-\left( F'(R)R-F(R)\right)+(4\dot{H}+6H^2)(F'(R)-1)
+4H F'(R)+2\ddot{F}'(R)
\Bigr]\, \nonumber\\
\end{eqnarray}
Here, the prime denotes the derivative with respect to $R$ and the dot represents the derivative with respect to cosmological time. Thus, we recover the Friedmann-like equations, and the modification to gravity has a fluid EoS in the form of Equation~(\ref{start}).
For example, we may take $\omega(\rho_\mathrm{F})=\omega$, where $\omega$ is the EoS parameter of standard matter, and identify
the effective bulk viscosity as:
\begin{eqnarray}
\hspace{-5mm}
B(\rho_{\mathrm{F}},a(t),H, \dot{H}...) &=&
\frac{1}{2\kappa^{2}} \biggl\{ (1+\omega)(F(R)-R F'(R))
+(F'(R)-1)
\left[6H^2(1+\omega)+4\dot{H}\right]
\nonumber \\
\hspace{-5mm}
& &
+ H{\dot{F}}'(R)(4+6\omega)
+2{\ddot{F}}'(R)\biggr\}\,
\end{eqnarray} 

An interesting example of viable inhomogeneous fluid has been proposed in~\cite{BigRipfluid}. It is worth noting that such fluid brings a realistic scenario of the Universe today, but provides a final evolution different from the one associated with the $\Lambda$CDM
 model. 
The EoS is given by:
\begin{equation}
p_{\mathrm{F}}=-\rho_{\mathrm{F}}+f(\rho_{\mathrm{F}})\,
\end{equation}
where:
\begin{eqnarray}
\left\{\begin{array}{lll}
f(\rho_{\mathrm{F}})&=&+\frac{2\rho_{\mathrm{F}}}{3n}\left(1-\frac{4n}{\delta}\left(\frac{3\tilde{m}^2}{\kappa^2\rho_{\mathrm{F}}}\right)^{\frac{1}{2}}\right)^{\frac{1}{2}}\,,\quad t \leq t_0\, \\
f(\rho_{\mathrm{F}})&=&- \frac{2\rho_{\mathrm{F}}}{3n}\left(1-\frac{4n}{\delta}\left(\frac{3\tilde{m}^2}{\kappa^2\rho_{\mathrm{F}}}\right)^{\frac{1}{2}}\right)^{\frac{1}{2}}\,,\quad t >t_0\,
\end{array}\right.
\end{eqnarray} 
In the above expressions, $n\geq 1$ and $\delta$ are constant positive parameters, $\tilde{m}^2$ is a mass scale related with the matter energy density today as $\rho_{\mathrm{m}(0)}=3\tilde{m}^2/\kappa^2$, and $t_0$ is a fixed time. Moreover, at $t=t_0$, the fluid energy density has a minimum and $f(\rho_{\mathrm F})=0$. 
The EoS parameter, $\omega(\rho_{\mathrm{F}})$, reads:
\begin{equation}
\omega(\rho_{\mathrm{F}})\equiv\frac{p_{\mathrm{F}}}{\rho_{\mathrm{F}}}=-1+\sigma(t)\frac{2}{3n}\left(1-\frac{4n}{\delta}\left(\frac{3\tilde{m}^2}{\kappa^2\rho_{\mathrm{F}}}\right)^{\frac{1}{2}}\right)^{\frac{1}{2}}\,
\end{equation}
where $\sigma(t)=1$, when $t\leq t_0$, and $\sigma(t)=-1$, when $t>t_0$. In fact, $t=t_0$ ($f(\rho_{\mathrm{F}})=0$) is the transition point between quintessence ($-1<\omega(\rho_{\mathrm{F}})<-1/3$) and phantom ($\omega_{\mathrm{F}}<-1$) regions, such that $\omega(\rho_{\mathrm{F}})=-1$. 
More specifically, for $t<t_0$, $-1<\omega(\rho_{\mathrm{F}})<-1+2/(3n)\leq-1/3$, and for $t>t_0$, $-5/3\leq-1-2/(3n)<\omega(\rho_{\mathrm{F}})<-1$. 
The present accelerated epoch can be set at the time, $t=t_0$. 
From the fluid energy conservation law:
\begin{equation}
\dot{\rho}_{\mathrm{F}}+3Hf(\rho_{\mathrm{F}})=0\,
\end{equation}
one has:
\phantom{line}
\begin{equation}
\rho_{\mathrm{F}}=\frac{3\tilde{m}^2\left(\frac{a(t)}{n}\right)^{\frac{2}{n}}\left(4n+c_0^{-\left(\frac{1}{2}\right)}\left(\frac{a(t)}{n}\right)^{-\frac{1}{n}}\right)^4c_0}{16\delta^2\kappa^2}\,\label{generalFBigRip}
\end{equation}
\phantom{line}\\
where $c_0>0$ is an integration constant. We put $a(t_0)=1$ and 
impose the ratio between fluid energy density [$\rho_{\mathrm{F}(0)}$] and matter today as
$\rho_{\mathrm{F}(0)}/\rho_{\mathrm{m}(0)}=\Lambda/(3\tilde{m}^2)$, $\Lambda$ being the Cosmological Constant.
From $\dot{\rho}_{\mathrm{F}(0)}=0$ (namely, $\omega(\rho_{\mathrm{F}(0)})=-1$), one obtains:\\
\phantom{line} 
\begin{equation}
\left\{\begin{array}{l}
c_0=\frac{1}{16}\left(n^{1-\frac{1}{n}}\right)^{-2}\, \\ \\ 
\frac{16n^2}{\delta^2}=\frac{\Lambda}{3\tilde{m}^2}\,\label{Ccondition}
\end{array}\right.
\end{equation}
\phantom{line}\\
when $t\ll t_0$, the matter energy density, $\rho_m\sim a(t)^{-3}$, grows up faster than the fluid energy density, and we recover the matter era. However, since at $t=t_0$, $\rho_{\mathrm{F(0)}}>\rho_{\mathrm{m(0)}}$, the fluid energy density overtakes the matter energy density in the recent past and an accelerated expansion takes place. 
The Friedmann equation, $3H^2/\kappa^2=\rho_{\mathrm{F}}$, and Equation (\ref{Ccondition}) lead to:
\begin{equation}
H(t)=\frac{n\left(\frac{\delta}{\sqrt{\tilde{m}^2}}\right)}{(t_s-t)\left(t-t_s+\frac{\delta}{\sqrt{\tilde{m}^2}}\right)}\,,\quad t<t_s\,\label{HBigRipfluid}
\end{equation}
where $t_s>0$ is a fixed time parameter and $\delta/\sqrt{\tilde{m}^2}>t_s$, in order to have $H(t)>0$ (expanding Universe).
As a consequence, the future Universe expansion becomes singular when $t$ approaches $t_s$ and
the Hubble parameter diverges at finite time (Big Rip). Hence, $t_s$ corresponds to the lifetime of the Universe. 
The fluid exits from the de Sitter phase evolving in a phantom region. Such a kind of realistic inhomogeneous fluid is compatible with the $\Lambda$CDM description today, but provides a different future~scenario.

Many fluids could bring the future Universe evolution to become singular. The simplest and \linebreak well-known case is represented by the phantom fluid with $p_{\mathrm{F}}=\omega_{\mathrm{F}}\rho_{\mathrm{F}}$ and $\omega_{\mathrm{F}}<-1$. If $\omega_{\mathrm F}$ is close to minus one, this kind of fluid describes a viable current acceleration. However, it admits a finite-future time singularity~\cite{Caldwell}, namely, the Big Rip. In fact, the Equation of State with the Friedmann equation, $3H^2/\kappa^2=\rho_{\mathrm{F}}$, lead to:
\begin{equation}
H(t)=-\frac{2}{3(1+\omega_{\mathrm{F}})}\frac{1}{(t_{0}-t)}
\end{equation}
where $t_0$ is the time at which the singularity occurs in the expanding Universe, being the Hubble parameter positively defined. 

Future finite-time singularities may also appear in the presence of bulk viscosity~\cite{Alessia, Brevik, Alessia(2),Sebastiani} (see also \cite{new1, new2, new3, new4, new5} for viscous cosmologies in General Relativity, string-driven inflation and singularities in higher-order gravity),
as in Equation~(\ref{start}). For example, if $\omega_{\mathrm{F}}$ is a constant and 
$B(\rho_{\mathrm{F}},a(t),H, \dot{H}...)=-(3H)^2\tau$, where $\tau$ is a constant
(as we will see, it means that the viscosity is proportional to the Hubble parameter), one has:
\begin{equation}
\dot{\rho_{\mathrm F}}+3H\rho_{\mathrm F}(1+\omega_{\mathrm{F}})= (3H)^{3}\tau\label{Hallo}\,
\end{equation}
Thus:
\begin{equation}
\rho_{\mathrm F}=\frac{27 h_0^{3}\tau}{(2+3 h_0(1+\omega_{\mathrm{F}}))(t_{0}-t)^{2}}\,\label{prova}
\end{equation}
is a solution with:
\begin{equation}
H=\frac{h_0}{(t_0-t)}\,
\end{equation}
and the Friedmann equation leads to the requirement, $h_0=2/[3\kappa^2\tau-3(1+\omega_{\mathrm{F}}))]$, with $[3\kappa^2\tau-3(1+\omega_{\mathrm{F}})]>0$~\cite{Brevik}. Also, in this case, one has the Big Rip at the finite time, $t_0$.

\section{Viscous Fluids Coupled with Dark Matter}

In this Section, in an attempt to solve the coincidence problem~\cite{Odintsovcouplingfluids,Viscousfluids}, we consider the possibility of 
coupling viscous fluids with dark matter (DM).
In standard cosmology, the energy density of (dark) matter decreases with the scale factor as $\rho_{\mathrm{DM}}=\rho_{\mathrm{DM(0)}}a(t)^{-3}$, and why we observe dark matter today and dark energy almost equal in amount is an open question. However, if we introduce a coupling between dark fluid and dark matter, we will see that when fluid becomes dominant in the dynamics of an FRW Universe, the de Sitter solution can be also realized 
with a constant energy contribution of fluid and dark matter. The ratio between dark energy fluid and matter depends on the coupling constant and can be set equal to the observed value.
Furthermore, if the de Sitter solution is stable, we also may avoid any singular future~scenario.

As the first step, we assume $p_{\mathrm{DM}}=0$. As a result, the conservation laws of fluid and dark matter in FRW space-time is:
\begin{equation}
\dot \rho_{\mathrm{F}} +3H(\rho_{\mathrm{F}}+p_{\mathrm{F}})=-Q_0\rho_{\mathrm{F}}\,\label{ECL1}
\end{equation}
\begin{equation}
\dot \rho_{\mathrm{DM}} +3H\rho_{\mathrm{DM}}=Q_0\rho_{\mathrm{F}}\,\label{ECL2}
\end{equation}
Here, $Q_0$ is the coupling constant, $\rho_{\mathrm{DM}}$ is the energy density of dark matter and $\rho_{\mathrm{F}}$ and $p_{\mathrm{F}}$ are the energy density and pressure of a viscous fluid. We consider the following form of fluid EoS:
\begin{equation}
p_{\mathrm{F}}=\omega(\rho_{\mathrm{F}})\rho_{\mathrm{F}}-3 H\zeta(H)\,\label{eq.state}
\end{equation}
where $\zeta(H)$ is the bulk viscosity, and in our ansatz, it depends only on the Hubble parameter, $H(t)$. In general, also, the EoS parameter of fluid, $\omega(\rho_\mathrm{F})$, is not a constant and may depend on the energy density.
On thermodynamical grounds, in order to obtain the positive sign of the entropy change in an irreversible process, $\zeta(H)$ has to be positive~\cite{Alessia, Brevik,Alessia(2)}. The stress-energy tensor of fluid turns out to be:
\begin{equation}
T_{\mu\nu}^{\mathrm{(fluid)}}=\rho_{\mathrm{F}} u_{\mu}u_{\nu}+\left(\omega(\rho_{\mathrm{F}})\rho-3H\zeta(H)\right)(g_{\mu\nu}+u_{\mu}u_{\nu})\,
\end{equation}
where $u_{\mu}=(1,0,0,0)$ is the four velocity vector. 
The FRW-equations of motion read:
\begin{equation}
\rho_{\mathrm{F}}+\rho_{\mathrm{DM}}=\frac{3}{\kappa^2}H^{2}\,,
\quad
p_{\mathrm{F}}=-\frac{1}{\kappa^2}\left(2\dot{H}+3H^2\right)\label{EOM1fluidDM}\,
\end{equation}
In what follows, we will analyze two different cases, namely $\omega(\rho_\mathrm{F})$, constant, and $\omega(\rho_\mathrm{F})$, not a constant.

\subsection{$\omega(\rho_{\mathrm{F}})$ Constant}

 Suppose we have the $\omega(\rho_{\mathrm{F}})=\omega_{\mathrm{F}}$ constant and bulk viscosity in the form:
\begin{equation}
\zeta(H)=\tau (3H)^n\,\label{v}
\end{equation}
with $\tau>0$ and $n$ being constants. 
The solution of Equation~(\ref{ECL1}) is:\\
\phantom{space}
\begin{equation}
\rho_{\mathrm{F}}=\rho_{\mathrm{F(0)}}\frac{\mathrm{e}^{-Q_0 t-3\omega_{\mathrm{F}} \log a(t)}}{a(t)^3}+\frac{\tau 3^{2+n} \mathrm{e}^{-Q_0 t-3\omega_{\mathrm{F}} \log a(t)}}{a(t)^3}\int\mathrm{e}^{Q_0 t'+3\omega_{\mathrm{F}} \log a(t')}a(t')\dot{a}(t')^2\left(\frac{\dot{a}(t')}{a(t')}\right)^n dt'\,\label{unobis}
\end{equation}
\phantom{space}\\
where $\rho_{\mathrm{F(0)}}$ is a positive integration constant.
The de Sitter solution is obtained by the choice, $H=H_{\mathrm{dS}}$, where the Hubble parameter corresponds to the present value of the accelerated Universe, and one has:
\begin{equation}
\rho_{\mathrm{F}}=\rho_{\mathrm{F(0)}}\mathrm{e}^{-t(Q_0+3H_{\mathrm{dS}}(1+\omega_{\mathrm{F}}))}+\frac{(3H_{\mathrm{dS}})^{n+2} \tau}{(Q_0+3H_{\mathrm{dS}}(1+\omega_{\mathrm{F}}))}\,,\quad\omega_{\mathrm{F}}\neq -\left(\frac{Q_0}{3H_{\mathrm{dS}}}+1\right)\,\label{duebis}
\end{equation}
As a consequence, the solution of Equation~(\ref{ECL2}) for dark matter reads:
\begin{equation}
\rho_{\mathrm{DM}}=\rho_{\mathrm{DM(0)}}\mathrm{e}^{-3H_{\mathrm{dS}}t}-\rho_{F(0)}\frac{Q_0}{Q_0+3H_{\mathrm{dS}}\omega_{\mathrm{F}}}\mathrm{e}^{-t(Q_0+3H_{\mathrm{dS}}(1+\omega_{\mathrm{F}}))}+\frac{(3H_{\mathrm{dS}})^{n+1}Q_0\tau}{(Q_0+3H_{\mathrm{dS}}(1+\omega_{\mathrm{F}}))}\,\label{trebis}
\end{equation}
where $\rho_{\mathrm{DM(0)}}$ is a positive constant.
It is important to note that when dark matter is dominant, we can neglect the contribution of fluid in the matter EoS and 
$\rho_{\mathrm{DM}}\simeq\rho_{\mathrm{DM(0)}}a(t)^{-3}$, such that we recover the standard cosmology in the matter era~\cite{Viscousfluids}.
However, on the de Sitter solution, if $\tau\neq 0$, the EOMs [{Equation}~(\ref{EOM1fluidDM})] are satisfied only by putting $\rho_{\mathrm{F(0)}}=\rho_{\mathrm{DM(0)}}=0$. Therefore, we require:
\begin{equation}
\frac{\rho_{\mathrm{DM}}}{\rho_{\mathrm{F}}}=\frac{Q_0}{3H_{\mathrm{dS}}}=\frac{1}{3}\,
\end{equation}
and the coincidence problem may be solved by setting:
\begin{equation}
Q_0=H_{\mathrm{dS}}\,\label{Qcoupling} 
\end{equation}
The ratio of DM and fluid is approximately $1/3$, almost independent from initial conditions. From the second EOM of~ {Equation}~(\ref{EOM1fluidDM}), one derives the relation between $\omega_{\mathrm{F}}$ and $\tau$: 
\begin{equation}
 \omega_{\mathrm{F}}=-\frac{4}{3}+4\kappa^2 (3H_{\mathrm{dS}})^{n-1}\tau\,\label{omegafluidDM}
\end{equation}
Here, Equation~(\ref{Qcoupling}) has been used. In this way, the fluid energy density~[{Equation}~(\ref{duebis})] turns out to be positive. Furthermore, since $\tau>0$, we must require $\omega_{\mathrm{F}}>-4/3$.
For example, a viscous fluid with $\omega_{\mathrm{F}}=-1$ possesses the de Sitter solution, $H_{\mathrm{dS}}$, if its bulk viscosity is: 
\begin{equation*}
\zeta(H)=\frac{(3H)^{n}}{12\kappa^2 (3H_{\mathrm{dS}})^{n-1}}\,
\end{equation*}
and the coupling constant with DM is given by $Q_0=H_{\mathrm{dS}}$.

If $\tau=0$ (non-viscous case), it is easy to see that Equations~(\ref{duebis}) and (\ref{trebis}) are de Sitter solutions of the EOMs only if $Q_0=-3(1+\omega_{\mathrm{F}})H_{\mathrm{dS}}$ and $\rho_{\mathrm{DM(0)}}=0$, such that the coincidence problem is solved by requiring~\cite{Odintsovcouplingfluids}:
\begin{equation}
\frac{\rho_{\mathrm{DM}}}{\rho_{\mathrm{F}}}=-(1+\omega_{\mathrm{F}})= \frac{1}{3}\,
\end{equation}
which leads to the phantom fluid:
\begin{equation}
\omega_{\mathrm{F}}=-\frac{4}{3}\, \label{000}
\end{equation}

Let us come back to the general case of $\tau\neq 0$. In order to investigate if the de Sitter solution is an attractor or not, we write the perturbation as:
\begin{equation}
H(t)=H_{\mathrm{dS}}+\Delta(t)\,\label{perturbazione}
\end{equation}
Here, $\Delta(t)$ is a function of the cosmic time, $t$, and it is assumed to be small. The second EOM of {Equation} (\ref{EOM1fluidDM})~gives: \\
\phantom{line} 
\begin{equation}
2\dot{\Delta}(t)+6H_{\mathrm{dS}}\Delta(t)\simeq 3H_{\mathrm{dS}}(n+1)\Delta(t)\,
\end{equation}
Here, some remarks are in order. Since the perturbed Equation~(\ref{unobis}) results are implicit, we have 
used Equation~(\ref{duebis}) with $Q=H_\mathrm{dS}$ (in fact, we say that at first approximation
near the de Sitter solution, $\rho_{\mathrm F}\sim H_{\mathrm dS}^{n+1}$).
Furthermore, Equation~(\ref{omegafluidDM}) has been taken into account. By assuming $\Delta(t)=\mathrm{e}^{\lambda t}$, we~find:
\begin{equation}
 \lambda+3H_{\mathrm{dS}}-\frac{3}{2}H_{\mathrm{dS}}(n+1)\simeq 0\,
\end{equation}
that is:
\begin{equation}
 \lambda\simeq \frac{3}{2}H_{\mathrm{dS}}(n-1)\,
\end{equation}
We easily see that, if $n<1$, the de Sitter solution is stable and the coupling of viscous fluid and dark matter generates a stable accelerated Universe with a constant rate of DM and fluid energy, such that the future singular scenario is avoided.

\subsection{$\omega(\rho_{\mathrm{F}})$ Not a Constant}

Let us consider a more general case, when the EoS parameter, $\omega(\rho_{\mathrm{F}})$, of viscous fluid is not a constant. A simple example is given by:
\begin{equation}
\omega(\rho_{\mathrm{F}})=\left[A_0\rho_{\mathrm{F}}^{\alpha-1}-1\right]\,\label{cucucu}
\end{equation}
where:
$A_{0}$ and $\alpha$ are constant parameters. From energy conservation law Equation~(\ref{ECL1}), one has:
\begin{equation}
\dot\rho_{\mathrm{F}}+3H A_{0}\rho_{\mathrm{F}}^{\alpha}+Q_0\rho_{\mathrm{F}}=(3H)^{n+2}\tau\,
\end{equation}
We still suppose to deal with a bulk viscosity proportional to $H^{n}$ as in Equation~(\ref{v}), $\tau>0$ and $n$ being constants. When $\alpha\gg1$, for the de Sitter solution, $H=H_{\mathrm{dS}}$, we obtain:
\begin{equation}
\rho_{\mathrm{F}}\simeq \left(\frac{\tau(3H_{\mathrm{dS}})^{n+1}}{A_{0}}\right)^\frac{1}{\alpha}\,\label{duebisbis}
\end{equation}
and the energy density of dark matter reads:
\begin{equation}
 \rho_{\mathrm{DM}}\simeq\frac{Q_0}{3H_{\mathrm{dS}}}\rho_{\mathrm{F}}\,
\end{equation}
In order to solve the coincidence problem, we require $Q_0=H_{\mathrm{dS}}$, again. 
If the fluid drives the accelerated expansion of the Universe, it follows from the EOMs~[{Equation} (\ref{EOM1fluidDM})] that we must put: 
\begin{equation}
A_{0}\simeq \tau(3H_{\mathrm{dS}})^{n+1}\left(\frac{\kappa^2}{3H_{\mathrm{dS}}^{2}}\right)^\alpha\,\label{omegabis}
\end{equation}
and one has:
\begin{equation}
\omega(\rho_{\mathrm{F}})\simeq -1+3(3H_{\mathrm{dS}})^{n-1}\kappa^2\tau\,\label{sturmtruppen}
\end{equation}
By making a perturbation around the de Sitter solution as 
in Equation~(\ref{perturbazione}), the second EOM gives: \\
\phantom{line} 
\begin{equation}
2\dot{\Delta}(t)+6H_{\mathrm{dS}}\Delta(t)\simeq H_{\mathrm{dS}}\left(\frac{n+1}{\alpha}\right)\Delta(t)\,
\end{equation}
\phantom{line}\\
where we have used Equations~(\ref{duebisbis}) and (\ref{omegabis}). By assuming $\Delta(t)=\mathrm{e}^{\lambda t}$, we finally have:
\begin{equation}
 \lambda+3H_{\mathrm{dS}}-\frac{1}{2}H_{\mathrm{dS}}\left(\frac{n+1}{\alpha}\right)\simeq0\,
\end{equation}
namely:
\begin{equation}
 \lambda\simeq H_{\mathrm{dS}}\left(\frac{1}{2}\left(\frac{n+1}{\alpha}\right)-3\right)\,
\end{equation}
Then, if $(n+1)/\alpha<6$, the de Sitter solution is a final attractor of the system, and we avoid future time~singularities.

\section{Constant Coupling of Viscous Fluids with Dark Matter}

In the previous section, we have reviewed some results following~\cite{Odintsovcouplingfluids,Viscousfluids}. In this section, we will 
present a new result generalizing the simplest case of a constant coupling between fluid and dark matter. In this case, due to the coupling constant, it is still possible to keep constant the matter energy density in the de Sitter Universe, in order to have a solution of the coincidence 
problem.
Recall that the \linebreak FRW-conservation laws of fluid and dark matter can be written as: 
\begin{equation}
\dot \rho_{\mathrm{F}} +3H(\rho_{\mathrm{F}}+p_{\mathrm{F}})=-Q_0\,\label{ECL11}
\end{equation}
\begin{equation}
\dot \rho_{\mathrm{DM}} +3H\rho_{\mathrm{DM}}=Q_0\,\label{ECL22}
\end{equation}
$Q_0$ being the coupling constant between fluid and dark matter.
In what follows, we will separately analyze the non-viscous case and the viscous case.

\subsection{Non-Viscous Case}

Let us start by considering the perfect fluid with
$p_{\mathrm{F}}=\omega_{\mathrm{F}}\rho_{\mathrm F}$,
where $\omega_{\mathrm{F}}$ is a constant. From Equation~(\ref{ECL11}) one has: \\
\phantom{line}
\begin{equation}
\rho_{\mathrm{F}}=\rho_{\mathrm{F}(0)}a(t)^{-3(1+\omega_{\mathrm{F}})}-Q_0\,a(t)^{-3(1+\omega_{\mathrm{F}})}\int a(t')^{3(1+\omega_{\mathrm{F}})}d t'\,
\end{equation}
\phantom{line}\\
Making use of the de Sitter solution, $H=H_\mathrm{dS}$, one finds:
\begin{equation}
\rho_{\mathrm{F}}=-\frac{Q_0}{3H_{\mathrm{dS}}(1+\omega_{\mathrm{F}})}+\rho_{\mathrm{F}(0)}e^{-3H_{\mathrm{dS}}t(1+\omega_{\mathrm{F}})}\,,\quad\omega_\mathrm{F}\neq -1\,
\end{equation}
and for the matter density [{Equation} (\ref{ECL22})], we obtain:
\begin{equation}
\rho_{\mathrm{DM}}=\rho_{\mathrm{DM(0)}}e^{-3H_{\mathrm{dS}}t}+\frac{Q_0}{3H_{\mathrm{dS}}}\,\label{**}
\end{equation}
In the above expressions, $\rho_{\mathrm{F}(0)}$ and $\rho_{\mathrm{DM}(0)}$ are constants.
Thus, Equation~(\ref{EOM1fluidDM}) are satisfied for \linebreak$\rho_{\mathrm{F}(0)}=\rho_{\mathrm{DM}(0)}=0$ and:
\begin{equation}
Q_0=\frac{9H_{\mathrm{dS}}^3(1+\omega_{\mathrm{F}})}{\omega_{\mathrm{F}}\kappa^2}\,,\quad\omega_{\mathrm{F}}\neq 0\,
\end{equation}
The coincidence problem may be solved by the choice:
\begin{equation}
\frac{\rho_{\mathrm{DM}}}{\rho_{\mathrm{F}}}=-(1+\omega_{\mathrm{F}})= \frac{1}{3}\,
\end{equation}
which leads to the same condition of Equation~(\ref{000}); namely, we have a phantom fluid with $\omega_{\mathrm{F}}=-4/3$ and $Q_0=9H_{\mathrm{dS}}^3/4$.
Note that the fluid energy density turns out to be positive.

Furthermore, let us consider the inhomogeneous case in Equation~(\ref{cucucu}), namely, $\omega(\rho_{\mathrm{F}})=\left(A_0\rho_{\mathrm{F}}^{\alpha-1}-1\right)$,
$A_{0}$ and $\alpha$ constants. When $H=H_{\mathrm{dS}}$, the solution of fluid conservation law equation~reads:
\begin{equation}
\rho_{\mathrm F}=\left(-\frac{Q_0}{3A_0 H_{\mathrm{dS}}}\right)^{\frac{1}{\alpha}}\,
\end{equation}
 Since, on the de Sitter solution, the energy density of matter is given by Equation~(\ref{**}),
in order to satisfy the EOMs, we must require $\rho_{\mathrm{DM(0)}}=0$ and:
\begin{equation}
A_0=-\frac{Q_0}{H_\mathrm{dS}}\left(\frac{1}{3}\right)\left(-\frac{Q_0}{3H_{\mathrm{dS}}}+\frac{3 H_{\mathrm{dS}^2}}{\kappa^2}\right)^{-\alpha}\,
\end{equation}
Therefore, the coincidence problem is solved by setting:
\begin{equation}
Q_0=\frac{9H_{\mathrm{dS}}^3}{4\kappa^2}\,\label{trtr}
\end{equation}
such that $A_0=-(4)^{\alpha-1}(3H_{\mathrm{dS}})^{2(1-\alpha)}/(3(\kappa^2)^{1-\alpha})$,
and the energy density of fluid and matter are positive~quantities.

\subsection{Viscous Case}

Now, we introduce a non-zero viscosity in the fluid pressure as in Equation~(\ref{v}), namely,
\linebreak $\zeta(H)=\tau (3H)^n$,
with $\tau>0$ and $n$ constants. For the sake of simplicity, $\omega(\rho_{\mathrm{F}})=\omega_{\mathrm{F}}$ is assumed to be constant.
The fluid conservation law equation leads to: \\
\phantom{line}
\begin{equation}
\rho_{\mathrm{F}}=\rho_{\mathrm{F}}a(t)^{-3(1+\omega_{\mathrm{F}})}-
a(t)^{-3(1+\omega_{\mathrm{F}})}\int a(t')^{-2+3(1+\omega_{\mathrm{F}})}
\left[Q_0 a(t')^2-3^{n+2}\tau\dot{a}(t')^2\left(\frac{\dot a(t')}{a(t')}\right)^n\right] dt'\,
\label{integral}
\end{equation}
and for the de Sitter case, $H=H_{\mathrm{dS}}$, one gets:
\begin{equation}
\rho_{\mathrm{F}}=\frac{-Q_0+(3H_{\mathrm{dS}})^{n+2}\tau}{3H_{\mathrm{dS}}(1+\omega_{\mathrm{F}})}+\rho_{\mathrm{F(0)}}
\mathrm{e}^{-3H_{\mathrm{dS}}t(1+\omega_{\mathrm{F}})}\,,\quad\omega_{\mathrm{F}}\neq -1\,
\end{equation}
For the matter density, we have:
\begin{equation}
\rho_{\mathrm{DM}}=\frac{Q_0}{3H_{\mathrm{dS}}}+\rho_{\mathrm{DM(0)}}\mathrm{e}^{-3H_{\mathrm{dS}}t}\,
\end{equation}
Thus, the EOMs [{Equation} (\ref{EOM1fluidDM})] are satisfied if $\rho_{\mathrm{F(0)}}=\rho_{\mathrm{DM(0)}}=0$ and with the choice:
\begin{equation}
\tau=\frac{(3H_{\mathrm{dS}})^{-n-2}\left[-Q_0\kappa^2\omega_{\mathrm{F}}+9H_{\mathrm{dS}}^3(1+\omega_{\mathrm{F}})\right]}{\kappa^2}\,\label{ult}
\end{equation}
By imposing the ratio between dark matter and viscous fluid equal to $1/3$, we find,
again, the condition [{Equation} (\ref{trtr})], namely, $Q_0=9H_{\mathrm{dS}}^3/(4\kappa^2)$. As a consequence, Equation~(\ref{ult}) reads:
\begin{equation}
\tau=\frac{(3H_{\mathrm{dS}})^{1-n}(4+3\omega_{\mathrm{F}})}{12\kappa^2}\,
\end{equation}
and in order to have $\tau>0$, we must require $\omega_{\mathrm{F}}>-3/4$. 
Note that the energy density of fluid and dark matter again turns out positive.

If $\omega_{\mathrm{F}}=-1$, the solution of Equation~(\ref{integral}) for $H=H_{\mathrm{dS}}$ is:
\begin{equation}
\rho_{\mathrm{F}}=\rho_{\mathrm{F(0)}}+t\left[-Q_0+(3H_{\mathrm{dS}})^{2+n}\tau\right]\,
\end{equation}
which is a solution of the EOMs only if:
\begin{equation}
Q=(3H)^{2+n}\tau\,
\end{equation}
namely, $\rho_{\mathrm{F}}=\rho_{\mathrm{F(0)}}$, where $\rho_{\mathrm{F(0)}}$ is a constant energy density. In this case, $\rho_{\mathrm{DM}}=(3H_{\mathrm{dS}})^{n+1}\tau$ and Equation (\ref{EOM1fluidDM}) are satisfied for:
\begin{equation}
\tau=\frac{(3H_{\mathrm{dS}})^{-n-1}(3H_{\mathrm{dS}}^2-\rho_{\mathrm{F(0)}}\kappa^2)}{\kappa^2}\,
\end{equation}
Finally, the coincidence problem is solved by requiring:
\begin{equation}
\rho_{\mathrm{F(0)}}=\frac{9H_{\mathrm{dS}}^2}{4\kappa^2}\,
\end{equation}
such that $\tau=(3H_{\mathrm{dS}})^{1-n}/(12\kappa^2)$ and $Q_0=9H_{\mathrm{dS}}^3/4\kappa^2$.

\setcounter{equation}{0}
\section{Conclusions}

In this short review, we have revisited some aspects of inhomogeneous viscous fluids in a flat FRW Universe. In 
principle, any modification of gravity may be written in the form of such a kind of fluid. As a result, one may make use of the framework of 
General Relativity, namely, the Friedmann equations, and the analysis turns out simplified. A large number of inhomogeneous fluids is compatible 
with the observed current accelerated expansion of the Universe, but they may produce different future scenarios with respect to the stable de Sitter solution of the $\Lambda$CDM model. In particular, a finite-time future singularity, namely the Big Rip, could appear.
In the second part of the paper, we have made use of the conservation laws in which a coupling of fluid and dark matter was present. 
Two different possible couplings have been investigated. By a coupling of inhomogeneous viscous fluid with dark matter, the coincidence problem 
may be solved, and if the de Sitter solution is stable, one may avoid future singularities. In fact, the coupling between fluid and dark matter may change the behavior of dark matter in expanding the Universe when fluid becomes dominant in the Friedmann equations, rendering it constant. 
As a consequence, the ratio between dark matter and fluid is determined by the constant coupling, and it is independent of the initial conditions. 

Other studies of inhomogeneous viscous fluids and the dark energy problem have been presented in~\cite{uno,due,tre,quattro,sei,sette}.


\end{document}